\documentclass[twocolumn,trackchanges]{aastex701}

\newcommand{\bpic}{$\beta$ Pic}
\newcommand{\bpicb}{$\beta$ Pic b}
\newcommand{\planet}{$\beta$ Pic d}
\newcommand{\water}{H$_2$O}
\newcommand{\methane}{CH$_4$}

\usepackage{amsmath}
\usepackage{url}
\usepackage{xcolor}
\usepackage{graphicx}
\usepackage{comment}
\usepackage{wrapfig}
\usepackage{bm}
\usepackage{natbib}

\begin{document}

\title{Discovery of an Exterior Third Planet Orbiting $\beta$ Pictoris}

\author[0000-0002-9027-4456]{Aidan Gibbs}\email[show]{abgibbs@ucsd.edu}
\affiliation{Department of Astronomy \& Astrophysics, University of California, San Diego, La Jolla, CA 92093, USA}

\author[orcid=0000-0003-2233-4821]{Jean-Baptiste Ruffio}\email{jruffio@ucsd.edu}
\affiliation{Department of Astronomy \& Astrophysics, University of California, San Diego, La Jolla, CA 92093, USA}

\author[orcid=0000-0002-9799-2303]{Alexis Bidot}\email{abidot@stsci.edu}
\affiliation{Space Telescope Science Institute, Baltimore, MD 21218, USA}

\author[orcid=0000-0002-7129-3002]{Travis S. Barman}\email{barman@lpl.arizona.edu}
\affiliation{Lunar and Planetary Laboratory, The University of Arizona, 1629 E. Univ. Blvd, Tucson, AZ 85721, USA}

\author[orcid=0000-0001-5173-2947]{Clarissa R. Do \'O}\email{cdoo@caltech.edu}
\affiliation{Department of Astronomy, California Institute of Technology, Pasadena, CA 91125, USA}

\author[orcid=0000-0002-9936-6285]{Quinn M. Konopacky}\email{qkonopacky@ucsd.edu}
\affiliation{Department of Astronomy \& Astrophysics, University of California, San Diego, La Jolla, CA 92093, USA}

\author[orcid=0000-0002-3191-8151]{Marshall D. Perrin}\email{mperrin@stsci.edu}
\affiliation{Space Telescope Science Institute, Baltimore, MD 21218, USA}

\author[orcid=0000-0003-3708-241X]{Aneesh Baburaj}\email{ababuraj@northwestern.edu}
\affiliation{Center for Interdisciplinary Exploration and Research in Astrophysics, Evanston, IL 60201, USA}

\author[orcid=0009-0005-7704-5527]{Beck Dacus}\email{bdacus@ucsd.edu}
\affiliation{Department of Astronomy \& Astrophysics, University of California, San Diego, La Jolla, CA 92093, USA}

\author[orcid=0000-0003-1212-7538]{Bruce Macintosh}\email{bamacint@ucsc.edu}
\affiliation{Department of Astronomy and Astrophysics, University of California, Santa Cruz, Santa Cruz, CA 95064, USA}
\affiliation{University of California Observatories, 1156 High Street, Santa Cruz, CA 95064, USA}

\author[orcid=0000-0001-7443-6550]{Alex Madurowicz}\email{amadurowicz@stsci.edu}
\affiliation{Space Telescope Science Institute, Baltimore, MD 21218, USA}

\author[orcid=0000-0002-6618-1137]{Jerry W. Xuan}\email{jerryxuan@g.ucla.edu}
\altaffiliation{51 Pegasi b Fellow}
\affiliation{Department of Earth, Planetary, and Space Sciences, University of California, Los Angeles, CA 90095, USA}

\begin{abstract}

We report the discovery of $\beta$ Pictoris d (\planet), a third giant planet in the $\beta$ Pictoris system, which now becomes only the second directly imaged system with more than two confirmed planets. \planet\ was serendipitously detected in \textit{JWST}/NIRSpec IFU observations. A second epoch of NIRSpec and MIRI/MRS observations confirm the initial discovery. The extracted spectrum shows clear \methane, CO, and \water\ absorption features, and the planet's measured radial velocity is consistent with its orbital position. Radial velocity and astrometry measurements combined with orbital stability simulations suggest a semi-major axis $>30\,$au, consistent with \planet\ being responsible for carving the inner edge of the $\beta$ Pictoris debris disk. Using effective temperature estimates from atmosphere model grid fits combined with evolutionary models, we estimate a mass of $\sim2$--$4\,\text{M}_\text{Jup}$. \planet\ is the first planet discovered using spectral template matching with moderate-resolution spectroscopy, highlighting its sensitivity to planetary molecular features hidden within bright extrasolar debris disks that are difficult to access with broadband imaging.

\end{abstract}


\section{Introduction}\label{sec:intro}

Systems with multiple directly imaged planets are rare among the population of discovered exoplanets, with only a handful of systems known. These include HR 8799 b,c,d,e \citep{Marois2008_hr8799,Marois2010_hr8799e}, YSES 1 b,c \citep{Bohn_2020_Yses1b,Bohn_2020_Yses1c}, PDS 70 b,c \citep{Kepler2018_pds70b,Haffert2019_pds70c}, \bpic\ b,c \citep{Lagrange2009A&A...493L..21L,Lagrange2019NatAs...3.1135L}, HD 206893 B,c \citep{Milli2017A&A...597L...2M,Hinkley2023A&A...671L...5H}, and WISPIT-1 b,c \citep{vanCapelleveen2025A&A...704A.221V_wispit1}, of which only HR 8799 is confirmed to possess more than two imaged planets. While a much greater number of multi-planet systems have been discovered by transit and radial velocity methods, the majority of these planets are not directly observable due to their size or stellar separation. Imaged multi-planet systems are therefore uniquely valuable for understanding giant planet formation and evolution within a single natal disk. Recent studies of imaged multi-planet systems have explored compositional trends related to accretion history (e.g. \citealt{Nasedkin2024_hr8799,Zhang2024AJ....168..246Z_Yses1bc,Balmer_2025_hr8799,Ruffio2026arXiv260108227R, Xuan2026}), dynamical analysis to investigate formation physics, migration pathways, and disk interactions (e.g. \citealt{Wang2018AJ....156..192W,Zurlo2022_hr8799_dynamics, Lacquement2025A&A...694A.236L, Maas2025A&A...700A.108M_dynamics, Trevascus2025A&A...698A..19T_pds70, Do2025ApJ...995..190D_pds70}), and active accretion from circumplanetary disks (e.g. \citealt{Zhou2021AJ....161..244Z_pds70,Zhang2021Natur.595..370Z_Yses1b,Close2025AJ....169...35C_pds70,Hoch2025Natur.643..938H_Yses}). Most recently, \citet{Ruffio2026arXiv260108227R} and \citet{Xuan2026} demonstrated detection of sulfur and nitrogen species in the HR 8799 system, which can be used with measurements of carbon and oxygen abundances to examine planetary solid and gas accretion in the outer disk.

The $\beta$ Pictoris system (hereafter \bpic) is notable not only for its two previously discovered planets but also for hosting the first directly imaged debris disk \citep{Smith1984Sci...226.1421S}, which remains one of the most actively studied debris disks today. The disk displays a rich array of structural features including warps, misalignments, and asymmetries of the inner and outer disk (e.g. \citealt{Burrows1995AAS...187.3205B_innerdiskwarp, Mouillet1997,Heap2000ApJ...539..435H_innerdiskwarp,Golimowski2006AJ....131.3109G_innerdiskwarp,Apai2015ApJ...800..136A,Janson2021A&A...646A.132J_outerdisk, Rebollido2024AJ....167...69R,Lovell2026A&A...705A.200L_asymmetry}); gas and dust clumps (e.g. \citealt{Telesco2005Natur.433..133T_clumpassymetry,Li2012ApJ...759...81L_dustclump,Dent2014Sci...343.1490D_gasclump, Cataldi2018ApJ...861...72C_gasclump, Han2026arXiv260303540H}); and dynamically diverse populations of dust and planetesimals, including exocomets (e.g. \citealt{Pavlenko2022A&A...660A..49P,Vidal-Madjar1994A&A...290..245V_exocomets,Kiefer2014Natur.514..462K_exocomets,Matra2019AJ....157..135M}). Many of these structures have been linked to disk interactions with the two previously known giant planets, most prominently the inner disk warp \citep{Mouillet1997,Dawson2011ApJ...743L..17D_diskwarp}, but also the exocomet and dust populations \citep{Besut1996Icar..120..358B_exocomets,Beust2024A&A...683A..89B,Jaworska2026arXiv260305600J}. Indeed, the known planets alone cannot fully explain the disk dynamics and additional planets exterior to \bpicb\ have been hypothesized \citep{Freistetter2007A&A...466..389F,Matra2019AJ....157..135M,Skaf2023A&A...675A..35S_clumpdynamics,Lacquement2025A&A...694A.236L}. The possibility of recent major planetesimal and/or planetary collisions, or tidal disruption events of small planetesimals by larger planets, has also been frequently suggested based on disk morphology and variability \citep{Jackson2014MNRAS.440.3757J_collisions,Jones_2023_collisions,Han2023MNRAS.519.3257H_clumpmotion,Chen2024ApJ...973..139C_collisions,Avsar2024ApJ...975...40A_collisions,Rebollido2024AJ....167...69R}.

The \bpic\ planetary system has been extensively observed to measure the orbits, masses, and spectra of \bpicb\ and c, and also to set limits on the presence of additional planets. The deepest of these previous observations are joint ground-based VLT/SPHERE high-contrast and ESO 3.6m/HARPS radial velocity (RV) limits \citep{Lagrange2020A&A...642A..18L}, and coronagraphic \textit{JWST}/NIRCam photometry \citep{Kammerer2024AJ....168...51K}. These data suggested that there are no planets more massive than $\sim2$ or $\sim4\,\text{M}_\text{Jup}$, respectively, at separations of $\sim20\,$au within the disk plane. The NIRCam sensitivity within the disk plane is primarily limited by broadband disk brightness with considerably lower, sub-Jovian mass limits achieved outside the disk plane.

Here we present the discovery of a new imaged planet orbiting \bpic\ A.  The new planet, \planet, was serendipitously discovered in \textit{JWST}/NIRSpec integral field unit (IFU) observations intended to study the atmosphere of $\beta$ Pic b. These observations leverage spectral information at $\text{R}\sim2700$ to improve sensitivity to planets within the disk plane compared to NIRCam. The discovery was subsequently followed up with MIRI MRS IFU and additional NIRSpec IFU observations to improve temperature and orbital estimates. In Section \ref{sec:observations}, we summarize our observations of \bpic, including data reduction steps which are sensitive to gaseous planets $>0.''3$ from the star at the cost of continuum-subtracting the resulting spectra. Next, we describe in Section \ref{sec:results} the discovery of \planet, including initial constraints on the planetary properties and orbit. Finally, in Section \ref{sec:discussion} we discuss the significance of \planet\ for the \bpic\ debris disk, and for future exoplanet discovery efforts.

\smallskip
An independent and contemporaneous discovery of this planet using VLT/ERIS and archival observations is separately reported by Sutlieff et al. (2026, accepted).

\begin{figure*}[t]
    \centering
    \includegraphics[width=1.0\textwidth]{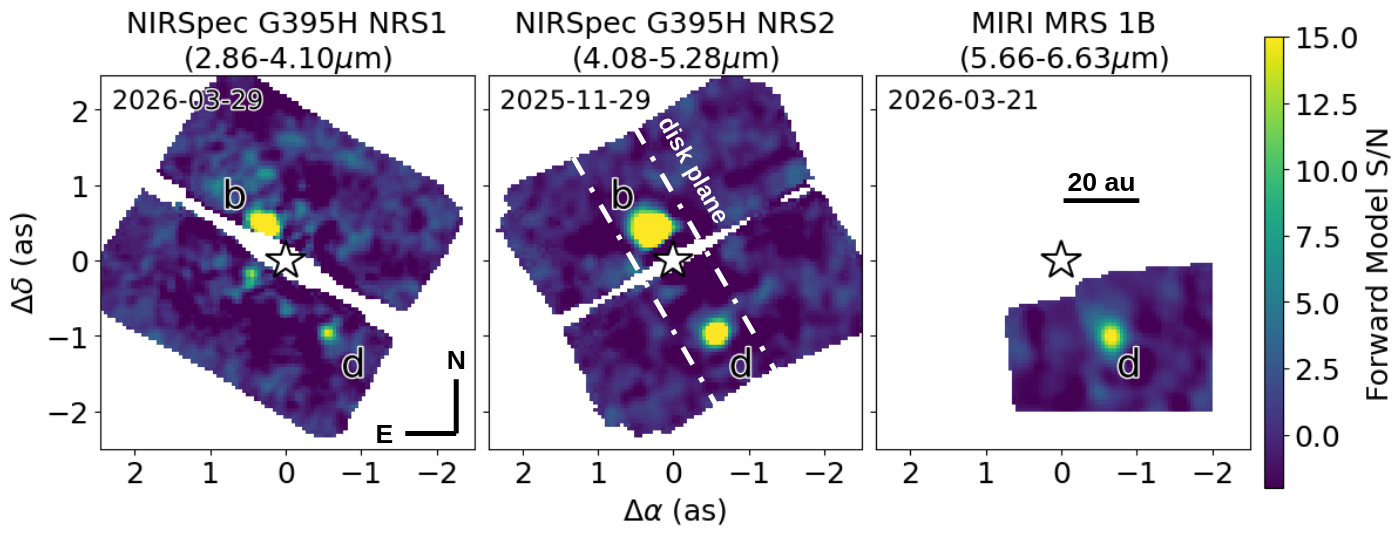}
    \caption{\textbf{Detection of \planet\ and b.} From left to right: S/N maps generated from the forward model framework (see Section \ref{sec:observations}) for the \textit{JWST}/NIRSpec IFU G395H/F290LP NRS1 and NRS2 detectors, on two different epochs, and the \textit{JWST}/MIRI MRS IFU channel 1B. Detections with G395H NRS1 and NRS2 are obtained in both epochs with similar significance, but are not shown for brevity. In both epochs of NIRSpec data, the detection significance in NRS1 is much less compared to NRS2, or to MRS 1B, due to the more limited number and depth of \methane\ lines compared to CO and \water. Both \planet\ and b lie in the disk plane, the orientation of which is indicated by the white dotted lines in the center map (offset to avoid overlap with planets). No significant movement of \planet\ relative to the star is measured between the two epochs, within a conservative astrometric precision of $\sim30\,$mas. Each NIRSpec image has a clipped region corresponding to the slices that are impacted by stellar saturation and charge transfer. The MIRI MRS data are taken with the star just outside the field of view, and therefore do not include \bpicb.}
    \label{fig:detection}
\end{figure*}

\section{Observations and Data Reduction}\label{sec:observations}

\bpic\ was observed with the \textit{JWST}/NIRSpec IFU \citep{Boker2023PASP..135c8001B} on 2025 November 28 (UT) as part of Cycle 4 GO program 8063 (PI: Jean-Baptiste Ruffio). Follow-up observations using the \textit{JWST}/MIRI MRS IFU \citep{Wright2023PASP..135d8003W} and NIRSpec IFU to confirm and improve characterization of the planetary candidate identified from the first observation epoch occurred on 2026 March 21 and 29 (UT) respectively (Cycle 4 DDT program 12518; PI: Aidan Gibbs). NIRSpec observations used the G395H/F290LP grating/filter covering $2.87$--$5.14\,\mu$m at $\text{R}\sim2700$. MIRI MRS observations used only the medium sub-band, covering $5.66$--$6.63$ and $8.67$--$10.13\,\mu$m in channel 1 and 2 at $R\sim2750$--$3750$. All \textit{JWST} data used in this paper can be found in MAST: \dataset[10.17909/sf76-2938]{http://dx.doi.org/10.17909/sf76-2938}.

\subsection{NIRSpec G395H} \label{sec:nirspec}

Each NIRSpec IFU epoch captured 32 total exposures across two observatory rolls separated by $\sim10^\circ$ for GO 8063 and $\sim4^\circ$ for DDT 12518 (16 exposures per roll). 
A small cycling dither ($\sim0.''25$) was applied between exposures to improve the spatial sampling of the NIRSpec IFU in reduced data products. All G395H/F290LP exposures used 4 groups per integration and 3 integrations per exposure, for a total exposure time of $5154\,$seconds per epoch. Because \bpic\ is too bright for target acquisition, pointing was established via guide-star acquisition, with an absolute pointing accuracy of $\sim0.''1$.

The uncalibrated NIRSpec IFU detector images were generated using version $2025\_4$a (SDP$\_$VER) of the JWST Science Data Processing subsystem. The data were reduced using the JWST data pipeline (version 1.20.2; \citealt{Bushouse_2025_17515973}) through stage 2 to produce 2D flux-calibrated images ($\_$\texttt{cal.fits}). The Calibration Reference Data System (CRDS) selection software version was 13.0.6 (CRDS$\_$VER) and the CRDS context version was jwst$\_$1464.pmap (CRDS$\_$CTX; \citealt{Greenfield2016A&C....16...41G}).

Additional data reduction steps were performed with the open-source python package \texttt{breads} \citep{agrawal_2024_11391503,Ruffio2024AJ....168...73R}. A detailed description of the reduction process can be found in \citet{Ruffio2024AJ....168...73R} and \citet{Ruffio2026arXiv260108227R}, and only an essential overview is presented here. No noteworthy changes are made to the reduction process as described in \citet{Ruffio2026arXiv260108227R}. 

To briefly summarize that reduction process, classical PSF subtraction is very challenging due to the spatial undersampling of the NIRSpec IFU and would not allow the detection of faint companions. The starlight is instead modeled row by row on the NIRSpec detectors using a continuum normalized stellar spectrum combined with a flexible spline model to reproduce the intensity variations of the PSF \citep{Ruffio2024AJ....168...73R}. The continuum normalized spectrum of \bpic\ A is derived empirically by normalizing every detector row using the spline method described in \citet{Ruffio2026arXiv260108227R} and then combined via a noise-weighted average. This stellar model is then used both for removing the starlight before extracting the spectrum of the companion and for forward modeling (FM) the data to compute the companion S/N maps.

Each IFU detector pixel maps to an R.A., declination, and $\lambda$, collectively referred to as the detector ``point-cloud.'' Because \bpic\ A is too bright for target acquisition, the World Coordinate System (WCS) headers carry a corresponding astrometric error. To determine the stellar centroid to $<0.''1$ precision, an \texttt{STPSF} PSF model for \textit{JWST} \citep{Perrin2014SPIE.9143E..3XP} is fit to the point cloud via weighted $\chi^2$ minimization outside of an inner working angle of $0.''3$.

To extract a spectrum at any physical coordinate in the IFU field with starlight removed, the stellar model was first subtracted across the full detector as an effective high-pass filter. Because all disk and planetary continuum is absorbed into the continuum spline during this fit, all resulting spectra are continuum-subtracted by this step. Furthermore, because the disk continuum is largely a combination of reflected starlight and thermal continuum at near-infrared wavelengths, the disk spectral contribution is also well subtracted. 
Although this step removes the continuum, the absolute flux calibration of the remaining high-pass filtered spectra is still retained (e.g. flux calibration of line depths). Next, for a given R.A. and declination centroid position, the PSF of a point source was fit to the point cloud, including all dithers, within an aperture of $0.''15$. This fit across wavelength returns the spectrum at the centroid position.

Extracting companion spectra requires knowledge of their positions. For companion detection, astrometry, and RV measurements, a forward model (FM) fit of the star, companion, and residuals is performed within the point cloud (see details in Section 4 of \citealt{Ruffio2024AJ....168...73R}).
The stellar component of the FM consists of the normalized stellar spectrum modulated by a spline, as described above, extended to include all detector rows whose traces overlap a given companion location. The companion model consists of an \texttt{STPSF} PSF whose flux is set by a user-provided template spectrum. Under the assumption that the companion's observed spectrum matches the template, the absolute companion flux can be estimated directly from the fit. The uncertainty on the fitted companion flux for a given centroid position defines the detection sensitivity, and the ratio of the inferred flux to uncertainty gives the detection S/N.

The detection S/N depends on how well the template spectrum matches the observed companion spectrum, as well as on the companion's position and RV. We evaluate the FM over a grid of RA, declination, and radial velocity in order to determine the presence of companions in the data, regardless of prior knowledge of their existence or positions. Because CO features remain present across a wide range of effective temperatures, this FM approach applied to G395H spectroscopy on the NRS2 detector from $4.1$--$5.3\,\mu$m is effective at detecting substellar companions across a broad range of properties without any fine tuning of the companion template.

\subsection{MIRI MRS Medium Channel} \label{sec:miri}

MRS data was obtained as part of Cycle 4 DDT 12518 and resulted in a total of 8 exposures in the medium sub-band.
In addition to the \bpic\ A observation, N Car was observed as a reference star to help calibrate spectral fringing that occurs as a result of reflective layers in the MIRI detector acting as an etalon \citep{Argyriou2020A&A...641A.150A,Gasman2025A&A...697A..58G}. 
The 2-point extended source dither pattern was selected as it was the best option to dither the companion in the field of view while keeping the offset host star PSF out of the field. This setup was motivated by known MRS straylight affecting the data when observing a bright star, nicknamed the zipper effect \citep{Argyriou2023A&A...675A.111A}. Because the dither patterns for MRS are limited to only 2 or 4 points, a 2x4 highly-overlapping mosaic was used to improve the spatial sampling of the MRS, in effect making 8 slightly-shifted copies of the 2-point pattern for a total of 16 distinct pointings. All exposures used 120 groups per integration to reduce the detector read out noise while avoiding saturation. The 4 integrations per exposure represented a total exposure time of 10720 seconds for each star. Both observations were centered using target acquisition so that the fringing pattern would be as similar as possible between the two observations of different stars.

Detector images were generated using the same SDP VER of the
JWST Science Data Processing subsystem and the same version of the JWST data pipeline as for NIRSpec/IFU.
The most significant difference in the data reduction process for MIRI MRS compared to NIRSpec is the special attention that is needed to carefully handle the fringing correction which is a significant systematic for MRS.  More details on this effect will be elaborated in a forthcoming analysis (Bidot et al. in prep). The default fringing correction was replaced with a customized fringing correction. The empirical fringe correction method described in \citep{Gasman2025A&A...697A..58G} was used to derive the fringe flat from the reference star observation. Unfortunately, the properties of the fringes are highly sensitive to the angles of incidence of the light projected onto the detector’s entrance. The fact that the disk is spatially resolved and bright, even in band 1B, has shown that the fringe pattern differs from the one that could be calibrated using the reference star. Therefore, we fitted a Fabry-Perot model directly to the \bpic\ data to mitigate the detector fringing. 

The \texttt{breads} framework described in the previous subsection was used to compute the S/N map resulting from the forward modeling and to extract a continuum-subtracted spectrum of companions. Both the pre-processing fringe correction and forward-modeling framework will be fully described in Bidot et al. (in prep).

\section{Results}\label{sec:results}

\begin{figure*}[t]
    \centering
    \includegraphics[width=1.0\textwidth]{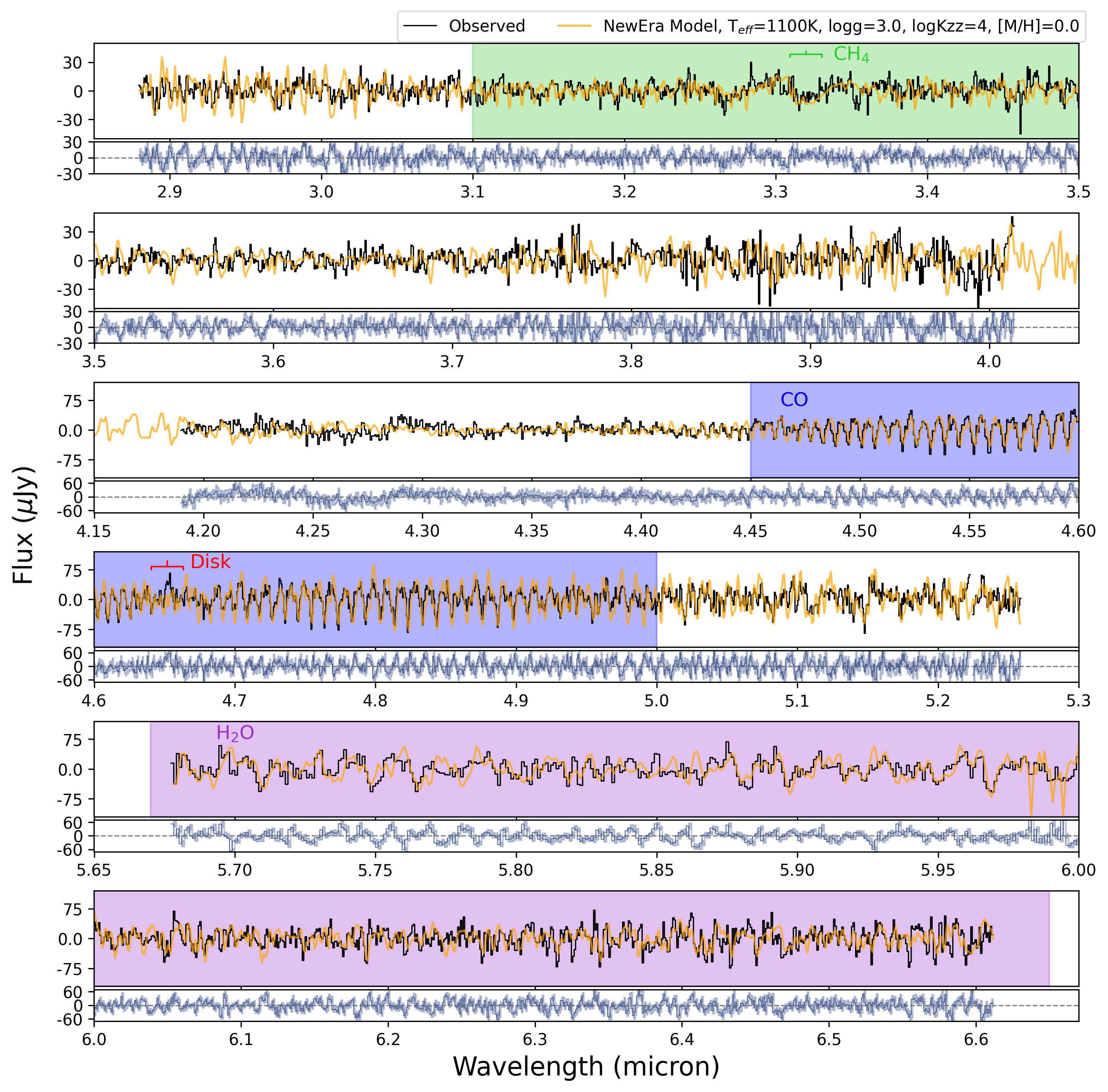}
    \caption{\textbf{Flux-calibrated \planet\ spectrum extracted from NIRSpec G395H/F290LP and MIRI MRS Channel 1B.} The observed spectrum (black) is continuum-subtracted as a consequence of the flexible continuum spline fit during the stellar subtraction process. A Phoenix NewEra model$^1$ is shown in orange for rough visual comparison. The model is continuum-subtracted using the same spline as the observations. Colored regions denote the dominant absorbing species for that wavelength region. Most noticeable are the prominent \methane\ feature at $3.3\,\mu$m and characteristic evenly spaced CO lines from $4.45$--$5.0\,\mu$m. Some wavelength regions of the spectrum are difficult to distinguish from noise (e.g. $3.8$--$4.3\,\mu$m), which only means that the amplitude of high frequency spectral features (lines) is fainter than the noise limit, not necessarily that the absolute flux of the planet is fainter. Only the high S/N detection of CO in the G395H NRS2 detector is used for forward modeling of \planet's astrometry and RV. The only noticeable spectral feature that is not planetary is a small disk emission feature of CO visible at $\sim4.65\,\mu$m, which is present in extracted spectra throughout the disk plane.}
    \label{fig:spectrum}
\end{figure*}

Figure \ref{fig:detection} shows the forward-model S/N maps for G395H/F290LP and MRS channel 1B combined across all exposures. The NIRSpec maps are produced using a BT-Settl (T$_\text{eff}=1100\,$K, $\log{g}=4.0$, solar composition; \citealt{Allard2003IAUS..211..325A}) spectral template, while the MIRI map is produced using an extrapolated retrieved spectrum for HR 8799 b \citep{Xuan2026}. These plots are only to illustrate detection and are too computationally expensive for property determination. Both epochs include observations with both detectors at similar significance, but only NRS1 and NRS2 are shown for the 2026 and 2025 epochs respectively for brevity and to illustrate the change in detector position and rotation. The template choice is not critically important for planet detection; \bpicb\ can be significantly detected with template effective temperatures ranging from at least $400$--$1600\,$K.

The S/N maps for NIRSpec show a strong detection of the previously known planet \bpicb\ $\sim0.''5$ north-east of the central star (which is subtracted in the maps), along with a previously unidentified point source $\sim1.1''$ ($\sim21.8\,$au projected) south-west of the star, which we identified as a planetary candidate in the first epoch. Both \bpicb\ and the candidate fall within the plane of the nearly edge-on ($i\sim89^\circ$) system. The detection of \bpicb\ and the candidate is more significant in G395H NRS2 ($4.08$--$5.28\,\mu$m) than NRS1 ($2.86$--$4.10\,\mu$m), owing to the increased contrast between \bpic\ A and both sources at shorter wavelengths, as well as the shallower line depths of \methane\ compared to CO. The detection significance of the candidate shown in Figure \ref{fig:detection} would only be slightly increased with a more fine-tuned template.

The candidate is also detected at the same stellar offset in MRS channel 1B \textbf{($5.66$--$6.63\,\mu$m)} observations. As the star is not observed in the MRS field, the offset calibration relies on the accuracy of the MRS target acquisition ($\sim90\,$mas). The candidate is not detected in channel 2B, primarily due to unsubtracted disk features at longer mid-infrared wavelengths that obscure it; however, detection may be possible in the future if these disk residuals can be modeled and subtracted, which is beyond the scope of this work. Because MRS IFU high-contrast observations require the star to be placed outside the detector field, \bpicb\ and the candidate cannot be detected simultaneously in a single observation, as they currently lie on opposite sides of the star. Since the candidate is detected at the same sky location with two independent instruments, and at 5 distinct observatory roll angles, it is unambiguously astrophysical. 

\setcounter{footnote}{0}

We extract the candidate spectrum at the location of the forward-model flux centroid (described in the following paragraph), shown in Figure \ref{fig:spectrum}. The peak spectral S/N is $\sim6$ per wavelength channel around $4.5\,\mu$m. Clear absorption features common to gaseous planets and brown dwarfs are visible. A $1100\,$K Phoenix NewEra Model\footnote{Higher temperature Phoenix NewEra models are described in \citet{Hauschildt2025A&A...698A..47H}; lower temperature models used here will be described in Barman et al. (in prep.).} is also shown for rough visual comparison. Notable features include \methane\ absorption at $3.3\,\mu$m, CO absorption between $\sim4.4$--$5.0\,\mu$m, and \water\ absorption at mid-infrared wavelengths. A background star or dust clump within the disk cannot produce these features. The only noticeable disk feature is isolated CO emission near $4.65\,\mu$m, identified in other extracted spectra from the disk plane.  Based on the spectrum alone, the candidate must be a gaseous planet, or less likely an unbound brown dwarf in the vicinity of \bpic. Note that some wavelength regions of the spectrum are challenging to distinguish from noise (e.g. $3.8$--$4.3\,\mu$m). This only means that the amplitude of high-frequency spectral features (i.e., lines) is near or below the noise limit at those wavelengths, not that the absolute flux is faint.

\subsection{Astrometry and Radial Velocity Measurements}\label{sec:astrometry}

\begin{deluxetable*}{lcc}\label{tab:astrometry}
\tablecaption{Astrometric and Radial Velocity Measurements of \planet}
\tablewidth{0pt}
\tablehead{
  \colhead{Parameter} &
  \colhead{2025-11-29} &
  \colhead{2026-03-29}
}
\startdata
RA (mas)  & $-528.0 \pm 10.7$ & $-525.5 \pm 30.8$ \\
Dec (mas) & $-1000.0 \pm 22.2$ & $-1001.7 \pm 22.7$ \\
Sep (mas) & $1132.2 \pm 20.1$ & $1132.5 \pm 26.7$ \\
PA (deg)  & $207.9 \pm 0.9$   & $207.7 \pm 1.7$ \\
RV (km/s) & $14.7^{+1.7}_{-2.3}$ & $12.6^{+2.4}_{-3.0}$ \\
\hline
$\Delta$RA (mas)  & \multicolumn{2}{c}{$2.5 \pm 32.6$}  \\
$\Delta$Dec (mas) & \multicolumn{2}{c}{$-1.7 \pm 31.8$} \\
$\Delta$Sep (mas) & \multicolumn{2}{c}{$0.3 \pm 33.4$}  \\
$\Delta$PA (deg)  & \multicolumn{2}{c}{$-0.2 \pm 1.9$}  \\
\enddata
\tablecomments{All dates in UT. Position angle (PA) is in degrees East of North. Radial velocities are relative to the solar system barycenter.}
\end{deluxetable*}

The two NIRSpec epochs give an astrometric baseline of $120\,$days to constrain proper motion of the candidate, as well as to measure the planet candidate's RV. All astrometric and RV fitting is performed using NIRSpec G395H NRS2 data since this wavelength region has the strongest detection. Furthermore, CO lines in the NRS2 wavelength range are very similar across a wide temperature range and we find that our choice of model template has no significant impact on astrometry and RV measurements. To determine the astrometric offset of the candidate from \bpic\ A, we again use the forward-model framework described in Sec. \ref{sec:observations}, using a $1100\,$K BT-Settl spectral template \citep{Allard2003IAUS..211..325A}. We narrow the forward-model region to a  $\pm0.''1$ window centered on the S/N maximum of the candidate source and \bpicb, sampled at $2\,$mas, using coarse positions measured across the full IFU field. We then fit a 2D Gaussian function to the forward-model flux map in each individual exposure using \texttt{photutils centroid\_2dg} method \citep{larry_bradley_2025_17129028}, weighted by the flux error map, to estimate the centroids and average them to find the combined centroid and uncertainty. We use the astrometry of \bpicb\ as a test of astrometric calibration error, as its orbital position is predicted from previous observations to within $3\,$mas \citep{Lacour2021}. The primary source of systematic astrometric uncertainty is from the fit to the stellar centroid due to its heavy saturation. From our astrometry of \bpicb, we identify a systematic astrometric bias of up to $\sim20\,$mas. The astrometric precision of NIRSpec IFU, which was not intended as a precision astrometric instrument, can be increased by improving stellar centroiding through the use of an empirical PSF (rather than a simulated \texttt{STPSF}), but these efforts are beyond the scope of this work.

Our measured astrometry of the candidate is presented in Table \ref{tab:astrometry}. We include the observed systematic bias of our \bpicb\ astrometry as an error term added in quadrature. For reference, a distant background star would be expected to have a $\Delta\alpha,\Delta\delta=-1.7, -27.8\,$mas due to stellar proper motion, and $67.8, -54.8\,$mas due to parallax between epochs. While our astrometry is not precise enough to confirm a bound planet using the stellar proper motion alone between our two epochs, we can constrain the candidate to within  $^{+16}_{-7}\,$pc of \bpic\ A. An object closer or farther than these limits would have moved beyond our $1\sigma$ limits between epochs due to relative parallax.

We further constrain whether the candidate is a bound planet by measuring its RV compared to \bpic\ A and b. We once again forward-model the same $1100\,$K spectral template with RV varying from $\pm100\,$km/s in $1\,$km/s steps, and fit a quadratic to the resulting probability distribution within $\pm10\,$km/s of its peak so that the best RV is not limited to the RV spacing. The derived radial velocities for the candidate are also presented in Table \ref{tab:astrometry}. Note that although the change in RV for a planet between our epochs is below our measurement precision, we nevertheless treat the epochs separately. Our RV uncertainty is estimated from the $68\%$ confidence interval of the forward-model posterior and independently verified with a bootstrap test. We verify our absolute RV calibration by measuring the RV of \bpicb\, which is consistent with the literature orbit within the uncertainty.

The RV of \bpic\ A is $20\pm0.7\,$km/s \citep{Gontcharov2006AstL...32..759G}, the measured RV of \bpicb\ is $\sim31\pm1\,$km/s, and the RV of the candidate is $\sim10$--$16\,$km/s, consistent with a planet orbiting \bpic\ A on the opposite side from \bpicb. The only possible false positive is a nearby L or T brown dwarf that is by chance in the line-of-sight of the \bpic\ disk and has an RV consistent with a planet. 
As a simple test of this possibility, we make an order-of-magnitude estimate of the probability of an unbound brown dwarf within $40\,$pc of Earth randomly lying within $<2''$ and $\pm10^\circ$ of the disk plane of \bpic\ A, assuming a high-end isotropic brown dwarf number density of $0.1\,$pc$^{-3}$ \citep{Kirkpatrick2024ApJS..271...55K}. Without considering the RV, this probability is already $<1:10^7$. We therefore conclude that the candidate is overwhelmingly likely to be bound, and confirm the planet, \bpic\ d.

\subsection{Effective Temperature and Mass Estimate}\label{sec:temp}

\begin{figure*}[t]
    \centering
    \includegraphics[width=0.75\textwidth]{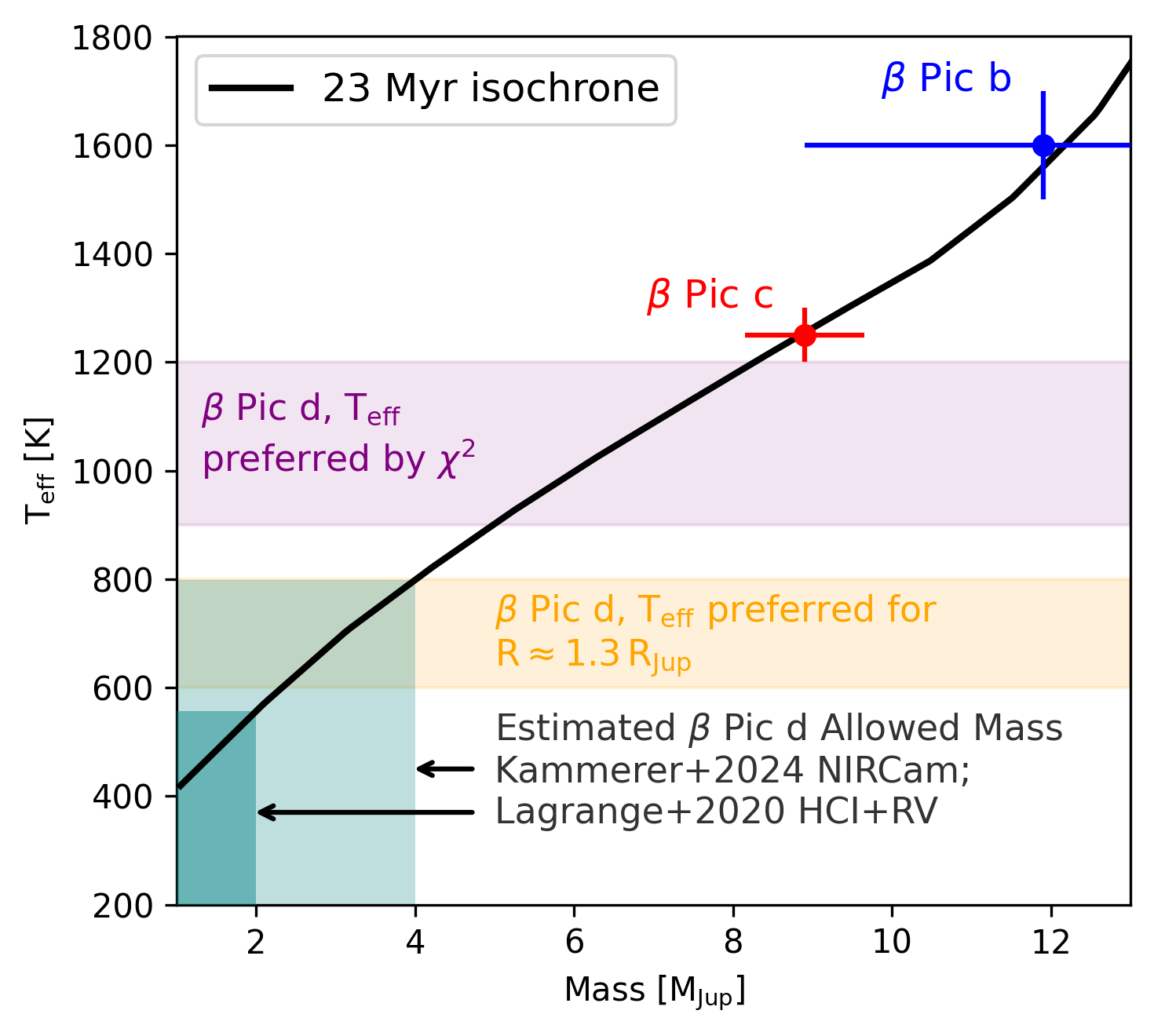}
    \caption{\textbf{Effective temperature versus mass isochrone interpolated for $23\,$Myr old planets from ATMO 2020 models \protect\citep{Phillips2020A&A...637A..38P}. } Literature masses and temperatures for \bpic\ b and c are shown for reference (temperatures aggregated from \protect\citealt{Nowak2020A&A...642L...2N,Kammerer2024AJ....168...51K,Ravet2025A&A...704A.325R}; masses from \protect\citealt{Lacour2021}). The range of best-fit T$_\text{eff}$ for \planet\ as determined by $\chi^2$ minimization to various grid models is shown as a purple shaded region. This shaded region does not represent a confidence interval, only the range of best-fits from different model families. An alternative T$_\text{eff}$ estimate using the relative line depth ratio of \methane\ to CO which is consistent with the data and the predicted evolutionary radius of $\sim1.3\,\text{R}_\text{Jup}$ is shown as an orange shaded region (see Section \ref{sec:temp} for more details). Our $\chi^2$ best-fit temperature estimates for \planet\ are in tension with previous mass upper limits for additional planets by \protect\citet{Lagrange2020A&A...642A..18L} and \protect\citet{Kammerer2024AJ....168...51K}, and with predicted evolutionary radii. We suggest that the true effective temperature is more likely in the orange shaded region, and that the limits from \protect\citet{Lagrange2020A&A...642A..18L} may be too low, since they would require a planetary radius much larger than expected.}
    \label{fig:evolution}
\end{figure*}

To constrain the effective temperature of \planet, we compare a series of atmospheric models to our continuum-subtracted planet spectra. This effective temperature estimate can provide an initial mass estimate for \planet\ via evolutionary cooling tracks in the absence of model-independent dynamical constraints. We compare our \planet\ spectra to Phoenix NewEra$^1$, Sonora ElfOwl \citep{Mukherjee2024ApJ...963...73M}, and Sonora Diamondback \citep{Morley_2024_12735103} models to sample a diverse range of models including disequilibrium chemistry and clouds. We do not explore atmospheric retrievals, which have been demonstrated for similar observations for example in \citet{Xuan2026}, due to their computationally expensive nature. For each model family, we first convolve and resample to the resolution and sampling of our observed spectra, then continuum-subtract the model using the same spline nodes and method as the data. We then calculate the multiplicative scale factor for the model needed to minimize the $\chi^2$, which can also be used to estimate the planet radius given the known distance to \bpic. 
The best-fit model depends on the model family; however, this method consistently favors effective temperatures in the $900$--$1200\,$K range. Increased metallicity models do not produce lower $\chi^2$ fits or favor lower temperatures. For cloudless models, low surface gravities $\log{\text{g}}\sim3.0$--$3.5$ are preferred while Sonora Diamondback models prefer higher gravities $\log{\text{g}\gtrsim4}$. The corresponding planetary radii for these fits are approximately Neptunian ($\sim0.3$--$0.4\,\text{R}_\text{Jup}$).  The results are similar even if the spectra are masked to exclude wavelengths dominated by noise. We caution that this temperature range is simply the most consistent with the shape of the continuum-subtracted spectra, but it is very unlikely to bound the true effective temperature based on the inferred radii, which is unphysical and far smaller than evolution models would predict \citep{Phillips2020A&A...637A..38P}. The most important influence on the continuum-subtracted spectral shape is the atmospheric chemistry, which is not only influenced in part by temperature, but also by composition and disequilibrium chemistry among other factors. The continuum-subtracted spectrum is therefore temperature degenerate. Only if a Neptune-like planet had experienced a recent major collision could these temperatures and radii be plausible.

As a simpler check of possible effective temperatures for \planet, we use our flux-calibrated line depths to estimate the observed line depth ratio between the $3.3\,\mu$m \methane\ feature and CO lines between $4.5$--$4.6\,\mu$m, and compare this to the same ratio calculated for models. Once again, this \methane/CO line depth ratio is highly degenerate with composition and disequilibrium chemistry, but provides a useful temperature diagnostic if those quantities are assumed independently. We can also use the multiplicative scale factor derived for the models (used to match the absolute line depths between data and model, rather than the ratio) to again calculate a model dependent radius for \planet. For solar metallicity and low mixing strength (log K$_{zz}=4$), a temperature range of $\sim900$--$1100\,$K can reproduce the \methane/CO line depth ratio with corresponding planetary radii in the range $\sim0.7$--$0.5\,R_\text{Jup}$. These radii are slightly larger than reported for the same temperatures in the previous paragraph because the optimal $\chi^2$ scaling always underscales the spectrum without a perfect fit. If metallicity and/or mixing strength are increased, colder models and larger radii are needed. For a [M/H]$=0.5$ and log K$_{zz}=7$, temperatures of $\sim500$--$700\,$K and corresponding radii of $\sim2$--$1.3\,R_\text{Jup}$ would be required to match the observed line depths. For reference, planets from $\sim1$--$8\,\text{M}_\text{Jup}$ are predicted to have radii of $\sim1.25$--$1.32\,R_\text{Jup}$ at the age of the \bpic\ moving group \citep{Lee2024MNRAS.528.4760L,Phillips2020A&A...637A..38P}. This analysis cannot meaningfully constrain surface gravity.

A $23\,$Myr isochrone of effective temperature versus mass from ATMO 2020 models \citep{Phillips2020A&A...637A..38P} is presented in Figure \ref{fig:evolution}. While \bpic\ b and c fall close to the isochrone, our estimated effective temperature from $\chi^2$ minimization is, as anticipated, hotter than expected based on previous upper limits on additional planets in the \bpic\ system from imaging and stellar RV measurements in \citet{Lagrange2020A&A...642A..18L} and \citet{Kammerer2024AJ....168...51K}. Our $\chi^2$ effective temperature would imply a mass of $\sim5$--$8\,$M$_\text{Jup}$, whereas those studies suggest additional planets more massive than $\sim2$--$4\,$M$_\text{Jup}$ would have been previously detected if they existed. Similar overestimated temperatures and underestimated radii have occurred with other imaged planets in the past (e.g. HR 8799; discussion in \citealt{Molliere2020A&A...640A.131M}).
Our temperature estimate from only the line depths, on the other hand, could be consistent with marginal non-detection by \citet{Kammerer2024AJ....168...51K} and \citet{Lagrange2020A&A...642A..18L} if the limits of \citet{Lagrange2020A&A...642A..18L} were slightly too constraining based, for example, on the complexity of accounting for disk subtraction, removing stellar pulsations in the RV data, or simply by inopportune timing if \planet\ was at smaller separation in the past decade. 
We suggest that the true effective temperature is most likely to be in the $\sim600$--$800\,$K range, because this range can produce radii consistent within $\sim20\%$ of evolutionary models, can match the absolute and relative line depths with fine-tuning of composition and mixing, and can be consistent with \citet{Kammerer2024AJ....168...51K} limits. This temperature range is imprecise based on the grid spacing of model spectra, so temperatures marginally outside this range might also be possible. These temperatures correspond to masses of $\sim2$--$4\,\text{M}_\text{Jup}$, rounded to the nearest whole Jupiter mass.

\subsection{Orbit and Stability}\label{sec:orbit}

\begin{figure*}[t]
    \centering
    \includegraphics[width=1.0\textwidth]{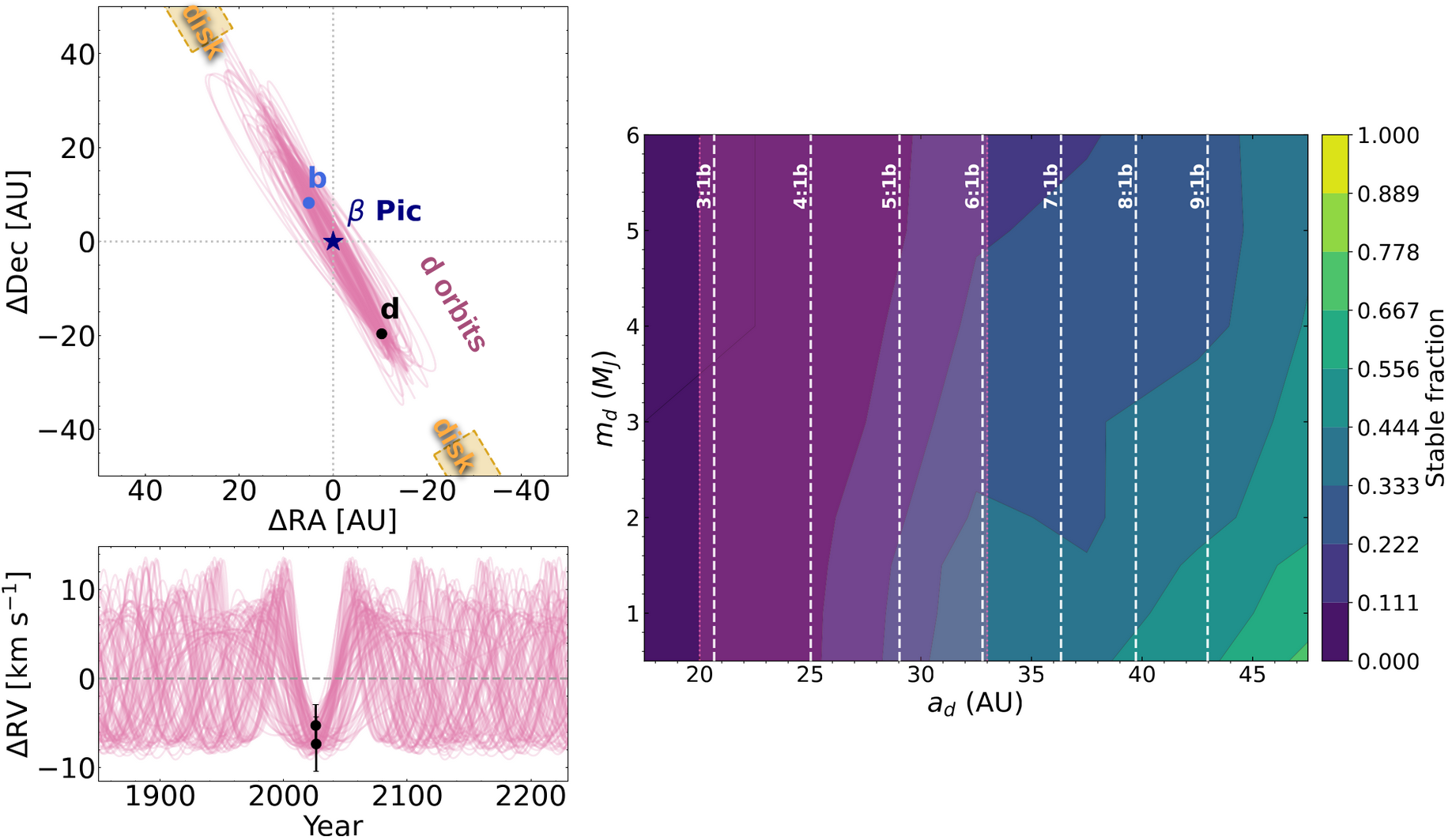}
    \caption{\textbf{\planet\ possible orbits and stability fractions. Left:} Allowed orbits (pink) from \texttt{Octofitter} \protect\citep{Thompson_2023} posteriors fit to astrometry and planetary RV (black) reported in Table \ref{tab:astrometry}, projected on-sky. Possible orbits are also shown in RV space. The locations of \bpicb\ and the approximate inner edge of the debris disk are plotted for reference. $\beta$ Pic c is not shown as it falls nearly coincident with the star. \textbf{Right:} Fraction of stable orbits as a function of \planet\ mass and semi-major axis, as determined by the MEGNO chaos indicator from orbit posteriors integrated over $5\,$Myr. The semi-major axis range of the $68\%$ percentile orbit posteriors is overlaid in pink. Orbital period ratios with \bpicb\ are shown as white dashed lines. The pink shaded region represents the 68th percentile of the orbital posteriors. The current astrometric baseline and RV precision do not strongly constrain the orbit; however, stability requirements suggest a semi-major axis of $>30\,$au.}
    \label{fig:orbit}
\end{figure*}

We fit for the orbit of \bpic\ d using \texttt{Octofitter} \citep{Thompson_2023} in a Bayesian framework, using the Pigeons Non-Reversible Parallel Tempering sampler \citep{Surjanovic2023}. We include the two epochs of astrometry and RV data relative to the star obtained in November 2025 and March 2026. Given the small amount of orbital coverage, which can bias the posteriors obtained in the orbit fit \citep{DoO2024}, we use prior information on the system to limit the range of some of the orbital parameters. Specifically, we limit the semi-major axis to 10--50 AU, such that it is at a wider semi-major axis than \bpic\ b ($\sim$ 10 AU) but inside the inner edge of the debris disk, which starts at $\sim$50 AU \citep{Lacquement2025A&A...694A.236L}. Priors on \bpic\ d's inclination and longitude of ascending node are set as truncated Gaussians with means $(i, \Omega) = (89^\circ, 31.5^\circ)$ and standard deviations of $5^\circ$ each, motivated by the system's disk orientation and the orbital solutions of \bpic\ b and c, which are coplanar with the disk to within ${\sim}1^\circ$ \citep{Lacour2021, Lacquement2025A&A...694A.236L}.
Finally, we also limit \bpic\ d's eccentricity to $<$0.4, motivated by dynamical stability and disk shaping results presented in \citealt{Lacquement2025A&A...694A.236L}. Our prior ranges and posterior draws are shown in Table \ref{tab:orbit_priors}. A selection of possible visual orbit solutions for \bpic\ d is shown in the left panel of Figure \ref{fig:orbit}.

\par
\par
\begin{deluxetable}{lll}
\tablecaption{Priors and posterior constraints for the \bpic\ d orbital fit. \label{tab:orbit_priors}}
\tablehead{
\colhead{Parameter} & \colhead{Prior} & \colhead{Posterior (median, $1\sigma$)}
}
\startdata
$a$ (AU)              & $\mathrm{LogUniform}(10, 50)$    & $24^{+8}_{-5}$ \\
$e$                   & $\mathrm{Uniform}(0, 0.4)$       & $0.19^{+0.14}_{-0.13}$ \\
$i$ (deg)             & $\mathcal{N}(89, 5)$             & $89 \pm 4$ \\
$\Omega$ (deg)        & $\mathcal{N}(31.5, 5)$           & $28.7^{+2.6}_{-1.7}$ \\
$\omega$ (deg)        & Uniform(0,360)                 & $191^{+127}_{-149}$ \\
$\tau$                & $\mathrm{Uniform}(0, 1)$         & $0.5 \pm 0.3$ \\
$P$ (yr)              & \nodata                          & $90^{+48}_{-24}$ \\
$M_\star$ ($M_\odot$) & $\mathcal{N}(1.75, 0.05)$        & $1.75 \pm 0.05$ \\
$\varpi$ (mas)        & $\mathcal{N}(51.44, 0.12)$       & $51.44 \pm 0.12$ \\
\enddata
\tablecomments{Posterior values quote the median and 68\% credible interval. All Gaussian priors are truncated at physically allowed bounds (e.g.\ $0 \leq i \leq 180^\circ$, $\varpi > 0$). We adopt the system parallax from \protect\citet{Nielsen2020} and a system mass consistent with the value used in the orbital analysis of \protect\citet{Lacour2021}. The orbital period $P$ is a derived quantity ($P = \sqrt{a^3/M_\star}$) and not directly sampled.}
\end{deluxetable}

In order to assess the stability of the system to further constrain \bpic\ d's semi-major axis and mass, we simulate the system's dynamical interactions using the N-body package REBOUND \citep{Rein2012}. The N-body integrator is WHFast \citep{Rein2015} with an integration timestep of 0.02 years. We use the orbital solutions and companion masses presented in \citealt{Lacquement2025A&A...694A.236L} for \bpic\ b and c, which come from the orbital determination in \citealt{Lacour2021}. Their work constrains the masses of \bpic\ b and c using precise GRAVITY astrometry and a Jacobi formulation that infers c's mass from b's orbital motion. Since the planets are not near a low-order mean-motion resonance, where changes in planet mass are most likely to affect stability, we therefore keep the masses and orbits of \bpic\ b and c fixed, varying only \bpic\ d's mass and orbital parameters. For \bpic\ d, we bin the orbital posterior samples by semi-major axis in 5 AU intervals across the prior range of 10--50 AU. We draw posteriors within those bins and evolve them for 5 Myr, corresponding to $\sim2 \times 10^5$ orbital periods of the dominant perturber \bpic\ b and therefore ample coverage of the secular interaction timescale. We evaluate stability for these binned posteriors for \bpic\ d mass values of 0.5, 1, 1.5, 2, 3, 4, 5, and 6 $M_{Jup}$ using the Mean Exponential Growth of Nearby Orbits (MEGNO) chaos indicator \citep{Cincotta2003}. We find that semi-major axis $>30$ AU is preferred for dynamically stable solutions, with a weak dependence on companion mass (see right panel of Figure \ref{fig:orbit}).

\section{Discussion}\label{sec:discussion}

The discovery of \planet\ has significant implications for interpretation of the \bpic\ system, including the debris disk. Previous studies predicted the possibility of one or two near- or sub-Jovian mass planets beyond the orbit of \bpicb\ based on the radius of the inner disk gap at $\sim50\,$au, nearly twice the radius predicted based on the inner two planets alone, and components of the disk morphology \citep{Augereau2001A&A...370..447A,Dent2014Sci...343.1490D_gasclump,Apai2015ApJ...800..136A,Ballering2016ApJ...823..108B,Cataldi2018ApJ...861...72C_gasclump,Matra2019AJ....157..135M,Lacquement2025A&A...694A.236L}. \planet\ could therefore be responsible for carving the inner disk edge. The outward-migration of a \planet\ like planet has also been proposed to explain both the disk asymmetry and the dynamically hot inclined disk population \citep{Matra2019AJ....157..135M}. It may also play a role in the distribution of infalling cometary bodies \citep{Beust2024A&A...683A..89B,Jaworska2026arXiv260305600J}, and could be responsible for the production of the south-west CO clump through a vortex trapping mechanism \citep{Skaf2023A&A...675A..35S_clumpdynamics, Rebollido2024AJ....167...69R}. The presence of \planet\ is likely a major validation of dynamical disk studies, but refinement of \planet's orbit and mass with future observations is required to determine whether it can explain much of the disk's unexplained dynamics, and whether a fourth massive planet interior to the disk is still possible.

\planet\ represents a novel discovery of a high-contrast exoplanet without the use of existing coronagraphic instruments. This unprecedented sensitivity is a direct result of moderate-resolution space-based spectroscopy, which allows effective subtraction of stellar and disk light that is not currently possible with broadband or low-resolution spectroscopic observations. Compared to prior NIRCam coronagraphy, spectroscopic detection with the NIRSpec IFU is largely immune to disk-scattered light, enabling an unambiguous and high S/N detection of \planet\ even in the middle of one of the brightest debris disks on the sky. Furthermore, the use of moderate-resolution spectroscopy allows high-confidence follow-up or confirmation of a planet from a single observation through detection of planetary absorption features and measurement of the planetary RV. Although secondary observations of \planet\ were performed prior to this publication, they only served as additional constraints on the orbit and temperature, and were not practically necessary for confirming the planet. The drawback of this spectroscopic discovery method is the loss of the planetary continuum, which for \planet\ makes accurate determination of the planetary effective temperature and mass difficult due to the degeneracy of line depths and shapes with temperature and other atmospheric physics, but may otherwise be sufficient for compositional studies \citep[e.g.][]{Xuan2022}. High-contrast detection and characterization of exoplanets with moderate spectral resolution IFUs is a promising method for future discovery of low-mass exoplanets that could be obscured by exozodiacal disks, and as an efficient method of simultaneous discovery and detection without the necessity of costly follow-up. This is particularly relevant for the design of the upcoming \textit{Habitable Worlds Observatory}.

\begin{acknowledgments}
This work is based on observations made with the NASA/ESA/CSA \textit{James Webb Space Telescope}. The data were obtained from the Mikulski Archive for Space Telescopes at the Space Telescope Science Institute, which is operated by the Association of Universities for Research in Astronomy, Inc., under NASA contract NAS 5-03127 for \textit{JWST}. These observations are associated with programs 8063 and 12518. 

Support for program 8063 was provided by NASA through a grant from the Space Telescope Science Institute, which is operated by the Association of Universities for Research in Astronomy, Inc., under NASA contract NAS 5-03127.

J.W.X. is grateful for support from the Heising-Simons Foundation 51 Pegasi b Fellowship (grant \#2025-5887).
\end{acknowledgments}

\begin{contribution}

A.G. led data analysis and manuscript preparation, including NIRSpec data reduction, astrometry and radial velocity extraction, and spectral analysis. A.G. is also PI of \textit{JWST} DDT 12518. J.B.R. is the PI of \textit{JWST} GO 8063 and Co-PI of \textit{JWST} DDT 12518, and provided project supervision. J.B.R. led development of the data reduction techniques included in the \texttt{breads} package which enabled this paper. J.B.R. helped design JWST DDT 12518 NIRSpec and MIRI observations. A.B. led \textit{JWST}/MIRI observations and data reduction, including designing \textit{JWST} DDT 12518 observations. T.B. developed additional Phoenix NewEra models for effective temperatures and chemistry relevant to giant planets, which were used in spectral analysis. C.D.O. led orbit fitting and stability analysis. Q.K. is Co-PI of \textit{JWST} GO 8063 and provided project supervision. M.P. helped design \textit{JWST} DDT 12518 NIRSpec and MIRI observations, and collaborated in the development of the data reduction techniques in \texttt{breads}. All authors contributed scientific expertise in data analysis, and proposal and manuscript preparation.

\end{contribution}

\facilities{JWST}

\software{\texttt{breads} \citep{agrawal_2024_11391503,Ruffio2024AJ....168...73R}; \texttt{photutils} \citep{larry_bradley_2025_17129028}; \texttt{octofitter} \citep{Thompson_2023}; \texttt{rebound} \citep{Rein2012}}

\bibliography{bibliography}{}
\bibliographystyle{aasjournalv7}

\end{document}